\documentstyle[aps]{revtex}

\begin{document}

\draft
\title{Entropy of a nonuniformly rectilinearly accelerating black hole} 
\author{He Han, Zhao Zheng}
\address{Department of Physics, Beijing Normal University, Beijing 100875, China}
\date{\today}
\maketitle

\begin{abstract}
Adopting thin film brick-wall model, we calculate the entropy of a
nonuniformly rectilinearly accelerating non-stationary black hole expressed by Kinnersley metric. 
Because the black hole is accelerated, the event horizon is axisymmetric.
The different points of horizon surface may have different temperature.
We calculate the temperature and the entropy density at every point of the horizon at first, 
then we obtain the total entropy through integration, which is proportional to
the aera of event horizon as the same as the stationary black holes. It is shown
that the black hole entropy 
may be regarded as the entropy of quantum fields just on the surface of event 
horizon.
\end{abstract}

\pacs{PACS number(s): 04.70.Dy}

\section{introduction}
Since Bekenstein suggested that the entropy of a black hole is proportional
to its surface area, the concerned research work has got much progress [1-3]. The 
brick-wall model suggested by 't Hooft gives a statistical explanation to the orgin 
of black hole entropy [4,5]. Recently, the brick-wall model is developed to the
thin film brick-wall model [6,7]. In this paper, we adopt the thin film brick-wall
model to calculate the entropy of a nonuniformly rectilinearly accelerating non-stationary black hole. 

Kinnersley has discussed the spacetime of an arbitrarily accelerating
point mass [8]. The metric used in this paper is a simple case in which the 
direction of acceleration doesn't vary, but the magnitude of acceleration
and the mass of black hole vary with time. 
Because the black hole is accelerated, the horizon is axisymmetric.
Unlike the spherically symmetric black hole, the different points of horizon
surface may have different temperature [9-12]. It is
obvious that in this spacetime the thermal equilibrium does not
exist in a large region, and the normal brick-wall method encounters difficulty
in this situation. The thin film brick-wall model is adopted to overcome this 
difficulty. 
Because the spacetime is axisymmetric and 
the different points of horizon surface may have different temperature,
we calculate the entropy density at every point of the horizon at first, 
then we obtain the total entropy through integration. 
If we appropriately select relationship between the thickness of film and its
distance to event horizon, the thickness vanishes when the distance goes to
zero, so the entropy of black hole can be regarded as the entropy of quantum
field just on the event horizon.

In Sec. II
we give the line element of spacetime and the surface equation of horizon. 
In Sec. III we use a method which
is proposed by Zhao and Dai to study the temperature of the horizon [9,10].
The entropy of black hole is calculated in Sec. IV in detail, and we give conclusion
and discussion in Sec. V.

\section{metric of nonuniformly rectilinearly accelerating black hole}

The metric of an arbitrarily accelerating point mass has been given by Kinnersley [8]
\begin{eqnarray}
ds^2&=&[1-2ar\cos\theta-r^2(f^2+h^2\sin^2\theta)-2Mr^{-1}]{\rm d}u^2+
2{\rm d}u{\rm d}r\nonumber\\
&&+2r^2f{\rm d}u{\rm d}\theta+2r^2h\sin^2\theta{\rm d}u{\rm d}\varphi-
r^2{\rm d}\theta^2-r^2\sin^2\theta{\rm d}\varphi^2,
\end{eqnarray}
where
\begin{eqnarray}
&f=-a(u)\sin\theta+b(u)\sin\varphi+c(u)\cos\varphi,\nonumber \\
&h=b(u)\cot\theta\cos\varphi-c(u)\cot\theta\sin\varphi. 
\end{eqnarray}
$a$,$b$,$c$ and $M$ are all arbitrary functions of retarted Eddington-Finkelstein
coordinate $u$. $a$
is the magnitude of acceleration. $b$ and $c$ describe the rate of change of the 
acceleration's direction. 
In the case of nonuniformly rectilinearly acceleration, we have $b=c=0$. The metric can be reduced to
\begin{eqnarray}
{\rm d}s^2&=&(1-2ar\cos\theta-r^2f^2-2Mr^{-1}){\rm d}u^2+2{\rm d}u{\rm d}r+
2r^2f{\rm d}u{\rm d}\theta\nonumber\\
&&-r^2{\rm d}\theta^2-r^2\sin^2\theta{\rm d}\varphi^2,
\end{eqnarray}
where
\begin{equation}
f=-a\sin\theta.
\end{equation}
Using the advanced Eddington-Finkelstein coordinate $v$ to replace the retarted
coordinate $u$, and adopting signature $(-,+,+,+)$, the metric becomes
\begin{eqnarray}
{\rm d}s^2&=&-(1-2ar\cos\theta-r^2a^2\sin^2\theta-2Mr^{-1}){\rm d}v^2+
2{\rm d}v{\rm d}r\nonumber\\
&&-2r^2a\sin\theta{\rm d}v{\rm d}\theta+r^2{\rm d}\theta^2+
r^2\sin^2\theta{\rm d}\varphi^2.
\label{original_metric}
\end{eqnarray}
$\theta=0$ is the direction of acceleration and the spacetime is axisymmetric. 
The determinant of metric is
\begin{equation}
g=-r^4\sin^2\theta,
\label{determinant}
\end{equation}
and the non-zero contravariant components of metric are
\begin{eqnarray}
g^{01}&=&g^{10}=1, \nonumber \\
g^{11}&=&1-2ar\cos\theta-2Mr^{-1}, \nonumber \\
g^{12}&=&g^{21}=a\sin\theta, \nonumber \\
g^{22}&=&r^{-2}, \nonumber \\
g^{33}&=&r^{-2}\sin^{-2}\theta.
\label{contravariant_components}
\end{eqnarray}

Now let us find the horizon equation of the spacetime represented by
Eq. (\ref{original_metric}). Considering the axial symmetry of spacetime, 
the surface equation of event horizon can be written as
\begin{equation}
H=H(v,r,\theta)=0
\ \ \ \ \mbox{or}\ \ \ \ 
r=r(v,\theta),
\label{surface_equation_a}
\end{equation}
which should satisfy null surface condition
\begin{equation}
g^{\mu\nu}\frac{\partial H}{\partial x^{\mu}}\frac{\partial H}
{\partial x^{\nu}}=0.
\label{null_surface_qualification}
\end{equation}
Substituting Eq. (\ref{contravariant_components}) into
Eq. (\ref{null_surface_qualification}), we get
\begin{equation}
2\frac{\partial H}{\partial v}\frac{\partial H}{\partial r}+\left( 1-2ar\cos\theta-\frac{2M}{r}\right)\left(\frac{\partial H}
{\partial r}\right)^2+2a\sin\theta\frac{\partial H}{\partial r}
\frac{\partial H}{\partial \theta}+\frac{1}{r^2}\left( \frac{\partial H}
{\partial \theta}\right)^2=0.
\label{surface_equation_b}
\end{equation}
From Eq. (\ref{surface_equation_a}) we can get
\begin{equation}
\frac{\partial H}{\partial r}\frac{\partial r}{\partial \theta}+
\frac{\partial H}{\partial \theta}=0,\ \ \ \ 
\frac{\partial H}{\partial r}\frac{\partial r}{\partial v}+
\frac{\partial H}{\partial v}=0.
\label{assistant_a}
\end{equation}
Substituting Eq. (\ref{assistant_a}) into Eq. (\ref{surface_equation_b}), we have
\begin{equation}
-2\dot{r}_H+\left(1-2ar_H\cos\theta-\frac{2M}{r_H}\right)-(2a\sin\theta)r_H'+\frac{{r_H'}^2}{r_H^2}=0,
\label{surface_equation_final}
\end{equation}
where
\begin{equation}
\dot{r}_H=\left( \frac{\partial r}{\partial v}\right)_{r=r_H},\ \ \ \ 
r_H'=\left( \frac{\partial r}{\partial \theta}\right)_{r=r_H}.
\end{equation}
Surface $r_H$ which satisfies Eq. (\ref{surface_equation_final}) is the event horizon of
nonuniformly rectilinearly accelerating black hole.

\section{temperature of black hole}

In this section, we use a method [9-11] which is proposed by Zhao and Dai to
study the temperature of the horizon. The calculation of this method is simple
and precision. It can be used to calculate temperature of various black holes, including non-spherically
symmetic and non-stationary black holes.

This method is based on the Damour-Ruffini's scheme [12]. The essential point
is that if we use tortoise coordinate, the radial part of Klein-Gordon equation
near the horizon will be reduced to the standard form of wave equation as
\begin{equation}
\frac{\partial^2\Phi}{\partial r_*^2}+2\frac{\partial^2\Phi}
{\partial v\partial r_*}=0.
\label{wave_equation}
\end{equation}

From Eq.(\ref{surface_equation_final}) we know that $r_H$ is the function of
$\theta$ and $v$ on the horizon. So the generalized tortoise coordinate
transformation is proposed as [10]
\begin{eqnarray}
r_*&=&r+\frac{1}{2\kappa}\ln [r-r_H(\theta,v)], \nonumber \\
v_*&=&v-v_0, \nonumber \\
\theta_*&=&\theta-\theta_0.
\label{tortoise_coordinate}
\end{eqnarray}
where $\kappa$ is an adjustable parameter. Both $v_0$ and $\theta_0$ are arbitrarily
fixed parameter. They are constant under the tortoise coordinate transformation. 

From Eq. (\ref{tortoise_coordinate}) we get
\begin{eqnarray}
\frac{\partial}{\partial r}&=&\left [1+\frac{1}{2\kappa(r-r_H)}\right]\frac{\partial}
{\partial r_*}, \nonumber \\
\frac{\partial}{\partial v}&=&\frac{\partial}{\partial v_*}-\frac{\dot{r}_H}{2\kappa(r-r_H)}
\frac{\partial}{\partial r_*},
\nonumber \\
\frac{\partial}{\partial\theta}&=&\frac{\partial}{\partial\theta_*}
-\frac{r_H'}{2\kappa(r-r_H)}\frac{\partial}{\partial r_*};
\label{partial}
\end{eqnarray}
Substituting Eq.(\ref{determinant}), Eq.(\ref{contravariant_components})
and Eq.(\ref{partial}) into Klein-Gordon equation
\begin{equation}
\frac{1}{\sqrt{-g}}\frac{\partial}{\partial x^\mu}
\left(\sqrt{-g}g^{\mu\nu}\frac{\partial \Phi}{\partial x^\nu}\right)
-\mu^2\Phi=0.
\label{Klein-Gordon}
\end{equation}

When $r \to r_H(v_0,\theta_0)$, $v \to v_0$ and $\theta \to \theta_0$, the 
equation can be reduced to
\begin{equation}
\alpha \frac{\partial^2\Phi}{\partial r_*^2}+2\frac{\partial^2\Phi}
{\partial r_* \partial v_*}+2\Omega \frac{\partial^2\Phi}{\partial r_* \partial \theta_*}
-G\frac{\partial \Phi}{\partial r_*}=0,
\label{reduced_KG_1}
\end{equation}
where
\begin{equation}
\alpha=\lim_{r \to r_H(v_0,\theta_0) \atop v \to v_0 , \theta \to \theta_0 }
\frac{
\{-2\dot{r}_H+(1-2ar\cos\theta-\frac{2M}{r})[2\kappa(r-r_H)+1]-(2a\sin\theta)r_H'\}
r^2[2\kappa(r-r_H)+1]+{r_H'}^2}
{2\kappa(r-r_H)[2\kappa(r-r_H)+1]r^2},
\end{equation}
\begin{equation}
\Omega=-\left( f+\frac{r_H'}{r_H^2} \right)_{v \to v_0 \atop \theta \to \theta_0},
\label{Omega1}
\end{equation}
\begin{equation}
G=\left( -\frac{2}{r_H}+16a\cos\theta+\frac{2M}{r_H^2}-\frac{{r_h'}^2}{r_H^3}
+\frac{r_H'}{r_H^2}\cot\theta \right)_{v \to v_0 \atop \theta \to \theta_0}.
\end{equation}
We select the adjustable parameter $\kappa$ as
\begin{equation}
\kappa=\left. \frac{1}{2r_H}\frac{\frac{M}{r_H^2}-a\cos\theta-\frac{{r_H'}^2}{r_H^3}}
{\frac{M}{r_H^2}+a\cos\theta+\frac{{r_H'}^2}{2r_H^3}} \right|_{v \to v_0 \atop \theta \to \theta_0}.
\label{kappa_general}
\end{equation}
Then we have $\alpha=1$, and Eq.(\ref{reduced_KG_1}) can be reduced to 
\begin{equation}
\frac{\partial^2\Phi}{\partial r_*^2}+2\frac{\partial^2\Phi}
{\partial r_* \partial v_*}+2\Omega\frac{\partial^2\Phi}{\partial r_* \partial \theta_*}
-G\frac{\partial \Phi}{\partial r_*}=0.
\label{reduced_KG_2}
\end{equation}
Separting variables as
\begin{equation}
\Phi=R(r_*)e^{-i \omega v_* + ik_\theta \theta_*+in\phi},
\end{equation}
we can verify that the radial wave solutions of Eq.(\ref{reduced_KG_2}) are, respectively,
\begin{eqnarray}
\phi_{\rm in}&=&e^{-i\omega v_*}, \\
\phi_{\rm out}&=&e^{-i\omega v_*+Gr_*+i(2\omega-2\Omega k_\theta) r_*}.
\end{eqnarray}
$\phi_{\rm in}$ is the ingoing wave, while $\phi_{\rm out}$ is the outgoing wave.
Near the event horizon $r_H$, $r_*=\frac{1}{2\kappa}\ln (r-r_H)$, $\phi_{\rm out}$ can be rewritten as
\begin{equation}
\psi_{\rm out}=e^{-i\omega v_*}(r-r_H)^{G/2\kappa}(r-r_H)^{i(\omega-\Omega k_\theta)/\kappa}.
\end{equation}
It is not analytical at the horizon. By analytical continuation rotating $-\pi$
through the lower-half complex {\it r}-plane, 
\begin{equation}
(r \to r_H) \to |r-r_H|e^{-i\pi}=(r_H-r)e^{-i\pi},
\end{equation}
we can extend $\phi_{\rm out}$ from
the outside of the black hole into the inside of the black hole [12]
\begin{equation}
\phi_{\rm out} \to \phi_{\rm out}'=e^{-i\omega v_*+Gr_*+i(2\omega-2\Omega k_\theta) r_*}e^{-i\pi G/2\kappa}
e^{\pi(\omega-\Omega k_\theta)/\kappa}
\end{equation}
The relative scattering probability of the outgoing wave at the horizon is
\begin{equation}
\left| \frac{\phi_{\rm out}}{\phi_{\rm out}'} \right|^2=e^{-2\pi(\omega-\Omega k_\theta) / \kappa}.
\end{equation}
Then the spectrum of the Hawking radiation is 
\begin{equation}
N_\omega=(e^{\frac{\omega-\Omega k_\theta}{T}}\pm 1)^{-1},
\label{Omega2}
\end{equation}
where 
\begin{equation}
T=\frac{\kappa}{2\pi}=\frac{1}{2\pi}\cdot \frac{1}{2r_H}\frac{\frac{M}{r_H^2}-a\cos\theta-\frac{{r_H'}^2}{r_H^3}}
{\frac{M}{r_H^2}+a\cos\theta+\frac{{r_H'}^2}{2r_H^3}}.
\label{temperature}
\end{equation}

We can see that the temperature of the nonuniformly rectilinearly acceleraing black hole
depends not only on the time, but also on the polar angle.

\section{Entropy of black hole}

In the brick-wall model proposed by 't Hooft [4], the black hole entropy is 
identified with the entropy of thermal gas of quantum fields excitations
outside the event horizon. This needs the thermal equilibrium between the
external fields and the black hole. However, this qualification is not satisfied
in the case of accelerating black hole, because the black hole is non-stationary 
and the temperature on the horizon is not
uniform as shown in Sec. III .

Recently, a new model, the thin film brick-wall model [6,7] is developed from the
original brick-wall model. This model considers that the entropy of
a black hole should be only related to quantum gas near its horizon.
Due to this opinion and the fact that
the density of quantum states near the horizon is divergent, it is natural
to take only the quantum field in a thin film near the event horizon into
accout. If we adopt the thin film brick-wall model, we find that although
the global thermal equilibrium is not satisfied, the local thermal equilibrium
is always exist. So the difficulty encountered by original brick-wall model is overcomed.

Because the spacetime is axisymmetric and the temperature on the
horizon is not uniform, we will calculate the entropy density at every point
of the horizon at first, then we obtain the total entropy through integration.

In the Sec. II, we have obtained the metric of nonuniformly rectilinearly
accelerating black hole shown by 
Eq.(\ref{original_metric}) and the horizon equation shown by Eq.(\ref{surface_equation_final}).
Introduce a coordinate transformation [13]
\begin{equation}
R=r-r_H(v,\theta),\ \ \ {\rm d}R={\rm d}r-\dot{r}_H{\rm d}v-r_H'{\rm d}\theta.
\end{equation}
The metric becomes
\begin{eqnarray}
{\rm d}s^2&=&-(-2\dot{r}_H+1-2ar\cos\theta-r^2a^2\sin^2\theta-2Mr^{-1}){\rm d}v^2+
2{\rm d}v{\rm d}R\nonumber\\
&&-2(r^2a\sin\theta-r_H'){\rm d}v{\rm d}\theta+r^2{\rm d}\theta^2+
r^2\sin^2\theta{\rm d}\varphi^2.
\label{base_metric}
\end{eqnarray}
The following calculation about entropy is based on this metric. 
The non-zero contravariant components of the metric are
\begin{eqnarray}
g^{01}&=&g^{10}=1, \nonumber \\
g^{11}&=&-2\dot{r}_H+1-2ar\cos\theta-2Mr^{-1}-2a\sin\theta r_H'+\frac{{r_H'}^2}{r^2}, \nonumber \\
g^{12}&=&g^{21}=a\sin\theta-\frac{r_H'}{r^2}, \nonumber \\
g^{22}&=&r^{-2}, \nonumber \\
g^{33}&=&r^{-2}\sin^{-2}\theta.
\label{contravariant_components_new}
\end{eqnarray}
$g^{11}=0$ is just the surface equation of horizon.

The form of metric can be changed to
\begin{eqnarray}
{\rm d}s^2&=&g_{00}{\rm d}v^2+2{\rm d}v{\rm d}R+2g_{02}{\rm d}v{\rm d}\theta
+g_{22}{\rm d}\theta^2+g_{33}{\rm d}\varphi^2 \nonumber\\
&=&\left( g_{00}-\frac{g_{02}^2}{g_{22}}\right){\rm d}v^2+2{\rm d}v{\rm d}R
+g_{22}\left( {\rm d}\theta+\frac{g_{02}}{g_{22}}{\rm d}v\right)^2+g_{33}
{\rm d}\varphi^2 \nonumber\\
&=&\hat{g}_{00}{\rm d}v^2+2{\rm d}v{\rm d}R
+g_{22}\left( {\rm d}\theta+\frac{g_{02}}{g_{22}}{\rm d}v\right)^2+g_{33}
{\rm d}\varphi^2,
\end{eqnarray}
where
\begin{equation}
\hat{g}_{00}=g_{00}-\frac{g_{02}^2}{g_{22}}=-
\left[-2\dot{r}_H+\left(1-2ar\cos\theta-\frac{2M}{r}\right)-(2a\sin\theta)r_H'+\frac{{r_H'}^2}{r^2}\right].
\end{equation}
We can get the surface equation of horizon from $\hat{g}_{00}=0$.
If we let
\begin{equation}
\frac{{\rm d}\theta}{{\rm d}v}=-\frac{g_{02}}{g_{22}},
\end{equation}
the metric can be changed to
\begin{equation}
{\rm d}s^2=\hat{g}_{00}{\rm d}v^2+2{\rm d}v{\rm d}R+g_{33}
{\rm d}\varphi^2.
\label{xianyuan}
\end{equation}
Like Kerr-Newman Black Hole, we can introduce a dragging angular velocity
\begin{equation}
\Omega=\frac{{\rm d}\theta}{{\rm d}v}=-\frac{g_{02}}{g_{22}}=a\sin\theta-\frac{r_H'}{r^2}.
\end{equation}
$\Omega$ is just that shown in Eq.(\ref{Omega1}) and Eq.(\ref{Omega2}).

Let us substitute the determinant and the contravariant components of metric
described by Eq. (\ref{base_metric})
into Klein-Gordon equation, which describes the scalar field with mass $\mu$,
\begin{equation}
\frac{1}{\sqrt{-g}}\frac{\partial}{\partial x^\mu}
\left(\sqrt{-g}g^{\mu\nu}\frac{\partial \Phi}{\partial x^\nu}\right)
=\mu^2\Phi.
\end{equation}
We suppose that the solution has the following form [7,14,15]
\begin{equation}
\Phi=e^{-i(Ev-m\varphi)}G(R,\theta),
\end{equation}
where
\begin{equation}
G(R,\theta)=e^{iS(R,\theta)}.
\end{equation}
With WKB approximation, we have
\begin{equation}
g^{11}k_R^2-2(E-g^{12}k_\theta)k_R+g^{22}k_\theta^2+m^2g^{33}+\mu^2=0,
\label{wkb}
\end{equation}
where
\begin{equation}
k_R=\frac{\partial S}{\partial R},\ \ \ \ k_\theta=\frac{\partial S}{\partial \theta}.
\end{equation}
From Eq. (\ref{wkb}) we can obtain the relationship between $k_R$ and $k_\theta$ as
\begin{equation}
{k_R}^+=\frac{E-g^{12}k_\theta}{g^{11}}+
\frac{\sqrt{(E-g^{12}k_\theta)^2-g^{11}(g^{22}k_\theta^2+m^2g^{33}+\mu^2)}}{g^{11}},
\end{equation}
\begin{equation}
{k_R}^-=\frac{E-g^{12}k_\theta}{g^{11}}-
\frac{\sqrt{(E-g^{12}k_\theta)^2-g^{11}(g^{22}k_\theta^2+m^2g^{33}+\mu^2)}}{g^{11}}.
\end{equation}

Free energy of the system is given by
\begin{equation}
F=-\int{\rm d}E\frac{\Gamma(E)}{e^{\beta (E-\Omega k_\theta)}-1},
\end{equation}
where $\Gamma(E)$ is the total number of modes whose energy is not greater than 
$E$. According to the semiclassical quantization condition and the thin film
brick-wall model, we have [4-7,14,15]
\begin{eqnarray}
\Gamma(E)&=&\frac{1}{4\pi^3}\int{\rm d}m\int{\rm d}\theta{\rm d}\varphi\int{\rm d}k_\theta
\left(\int_{\epsilon}^{\epsilon+\delta}k_R^+{\rm d}R
+\int^{\epsilon}_{\epsilon+\delta}k_R^-{\rm d}R\right)\nonumber\\
&=&\frac{1}{2\pi^3}\int{\rm d}m\int{\rm d}\theta{\rm d}\varphi\int{\rm d}k_\theta
\int_{\epsilon}^{\epsilon+\delta}\hat{k}_R{\rm d}R,
\end{eqnarray}
where
\begin{eqnarray}
\hat{k}_R&=&\frac{\sqrt{(E-g^{12}k_\theta)^2-g^{11}(g^{22}k_\theta^2+m^2g^{33}+\mu^2)}}{g^{11}} \nonumber \\
&=&\frac{\sqrt{(E-\Omega k_\theta)^2-g^{11}(g^{22}k_\theta^2+m^2g^{33}+\mu^2)}}{g^{11}}.
\label{kR}
\end{eqnarray}
Introducing $\tilde E=E-\Omega k_\theta$, we have
\begin{equation}
\hat{k}_R=\frac{\sqrt{{\tilde E}^2-g^{11}(g^{22}k_\theta^2+m^2g^{33}+\mu^2)}}{g^{11}},
\end{equation}
\begin{equation}
\Gamma(\tilde E)=\frac{1}{2\pi^3}\int{\rm d}m\int{\rm d}\theta{\rm d}\varphi\int{\rm d}k_\theta
\int_{\epsilon}^{\epsilon+\delta}\hat{k}_R{\rm d}R,
\label{Gamma}
\end{equation}
\begin{equation}
F=-\int_0^{+\infty}{\rm d}\tilde E\frac{\Gamma(\tilde E)}{e^{\beta \tilde E}-1}.
\end{equation}

The surface density of free energy on horizon can be expressed by
\begin{equation}
\sigma_F=-\int_0^{+\infty}{\rm d}\tilde E\frac{\sigma_\Gamma}{e^{\beta \tilde E}-1}.
\end{equation}
$\sigma_F$ and $\sigma_\Gamma$ are defined as
\begin{equation}
F=\int\sigma_F{\rm d}A,\ \ \Gamma=\int\sigma_\Gamma{\rm d}A,
\label{define}
\end{equation}
where
\begin{equation}
{\rm d}A=r_H^2\sin\theta{\rm d}\theta{\rm d}\varphi.
\end{equation}


Now, let us study the integration on $k_\theta$ and $m$ in Eq. (\ref{Gamma}). 
We use little mass approximation in the process of integration, and the result is
\begin{eqnarray}
\Gamma(\tilde E)&=&\frac{\tilde E^3}{6\pi^2}\int{\rm d}\theta{\rm d}\varphi
\int_{\epsilon}^{\epsilon+\delta}(g^{11})^{-2}(g^{22}g^{33})^{-\frac{1}{2}}{\rm d}R \nonumber\\
&=&\frac{\tilde E^3}{6\pi^2}\int{\rm d}\theta{\rm d}\varphi
\int_{\epsilon}^{\epsilon+\delta}(g^{11})^{-2}(g_{22}g_{33})^{\frac{1}{2}}{\rm d}R \nonumber\\
&\approx&\frac{\tilde E^3}{6\pi^2}\int {\rm d}A
\int_{\epsilon}^{\epsilon+\delta}(g^{11})^{-2}{\rm d}R \nonumber\\
&=&\int{\rm d}A\frac{\tilde E^3}{6\pi^2}\int_{\epsilon}^{\epsilon+\delta}(g^{11})^{-2}{\rm d}R.
\end{eqnarray}
Comparing with Eq. (\ref{define}), we get the surface density of $\Gamma(\tilde E)$ as
\begin{equation}
\sigma_\Gamma=\frac{\tilde E^3}{6\pi^2}\int_{\epsilon}^{\epsilon+\delta}(g^{11})^{-2}{\rm d}R.
\end{equation}
The surface density of free energy is given by
\begin{eqnarray}
\sigma_F&=&-\frac{1}{6\pi^2}\int_0^{+\infty}{\rm d}\tilde E\frac{\tilde E^3}{e^{\beta \tilde E}-1}
\int_{\epsilon}^{\epsilon+\delta}(g^{11})^{-2}{\rm d}R \nonumber \\
&=&-\frac{\pi^2}{90\beta^4}\int_{\epsilon}^{\epsilon+\delta}(g^{11})^{-2}{\rm d}R.
\end{eqnarray}
From
\begin{equation}
S=\left.\beta^2\frac{\partial F}{\partial\beta}\right|_{\beta=\beta_H},
\end{equation}
we can obtain the surface density of entropy, as
\begin{equation}
\sigma_S=\frac{4\pi^2}{90\beta_H^3}\int_{\epsilon}^{\epsilon+\delta}(g^{11})^{-2}{\rm d}R.
\label{entropy_a}
\end{equation}

Because $g^{11}=0$ is the surface equation of horizon, $g^{11}$
can be expressed by
\begin{equation}
g^{11}=f(r,\theta)(r-r_H).
\label{g11}
\end{equation}
Substituting Eq. (\ref{g11}) into Eq. (\ref{entropy_a}), we complete the integration on $R$, as
\begin{eqnarray}
\sigma_S&=&\frac{4\pi^2}{90\beta_H^3}\int_{\epsilon}^{\epsilon+\delta}
\frac{1}{f^2(r-r_H)^2}{\rm d}R \nonumber\\
&\approx&\frac{4\pi^2}{90\beta_H^3f_H^2}\int_{\epsilon}^{\epsilon+\delta}
\frac{1}{R^2}{\rm d}R \nonumber\\
&=&\frac{4\pi^2}{90\beta_H^3f_H^2}\frac{\delta}{\epsilon(\epsilon+\delta)}.
\label{entropy_b}
\end{eqnarray}
From Eq.(\ref{surface_equation_final}),Eq.(\ref{contravariant_components_new}) and Eq.(\ref{g11}), we can obtain the relationship between $f_H$ and $\kappa$, as
\begin{equation}
f_H=2\kappa(1-2\dot{r}_H-2ar_H'\sin\theta+\frac{2{r_H'}^2}{r_H^2}).
\label{kappa}
\end{equation}
Substituting Eq. (\ref{kappa}) into Eq. (\ref{entropy_b}), we get
\begin{equation}
\sigma_S=\frac{4\pi^2}{90\beta_H^3}\frac{1}{[\kappa(1-2\dot{r}_H-2ar_H'\sin\theta+\frac{2{r_H'}^2}{r_H^2})]^2}
\frac{\delta}{\epsilon(\epsilon+\delta)}\frac{1}{4}.
\label{shang3}
\end{equation}
Substituting $\beta_H=\frac{2\pi}{\kappa}$ into Eq. (\ref{shang3}), we get
\begin{eqnarray}
\sigma_S&=&\frac{1}{P}\frac{\delta}{\epsilon(\epsilon+\delta)}\frac{1}{4},
\end{eqnarray}
where
\begin{equation}
P=90\beta_H(1-2\dot{r}_H-2ar_H'\sin\theta+\frac{2{r_H'}^2}{r_H^2})^2.
\end{equation}

Selecting appropriate cut-off distance $\epsilon$ and thickness of thin film $\delta$
to satisfy
\begin{equation}
\frac{\delta}{\epsilon(\epsilon+\delta)}=P,
\label{relation}
\end{equation}
we can obtain the surface density of entropy, as
\begin{equation}
\sigma_S=\frac{1}{4}.
\label{density}
\end{equation}
The total entropy is certainly
\begin{equation}
S=\int\sigma_S{\rm d}A=\frac{1}{4}A_H.
\end{equation}

\section{conclusions}

By using the generalized tortoise transformation and the thin film brick-wall
model, we have studied the temperature and the entropy of 
a nonuniformly rectilinearly accelerating black hole. 
Because the black hole is accelerated, the horizon is axisymmetric and
the different points of horizon surface may have different temperature. 
To overcome the difficulty encountered
in normal brick-wall medol, we adopt the thin film brick-wall model in which
only the local thermal equilibrium is needed. We calculate the entropy density
at every point of the horizon at first, then we obtain the total entropy
through integration. The result is consistent with the well-known conclusion
that the black hole entropy is propotional to its area.

We believe that the black hole entropy should be just identified with the
entropy of the quantum field on the 2-dimensional surface of event
horizon [16]. The thin film
brick-wall model can easyly explain this opinion.
From Eq. (\ref{relation}) we can get
\begin{equation}
\delta=\frac{P\epsilon^2}{1-P\epsilon}.
\label{limitation}
\end{equation}
That is to say, if the relationship of the thickness $\delta$ of thin film and the distance
$\epsilon$ to horizon satisfies above equation, we can always get
Eq. (\ref{density}). Using Eq. (\ref{limitation}) we can get
\begin{equation}
\lim_{\epsilon \to 0}\frac{\delta}{\epsilon(\epsilon+\delta)}=
\lim_{\epsilon \to 0}\frac{\frac{P\epsilon^2}{1-P\epsilon}}
{\epsilon\left[\epsilon+\frac{P\epsilon^2}{1-P\epsilon}\right]}=
\lim_{\epsilon \to 0}\frac{\frac{P}{1-P\epsilon}}
{1+\frac{P\epsilon}{1-P\epsilon}}=P.
\end{equation}
This indicates that if appropriate relationship of thickness $\delta$
and distance $\epsilon$ is selected, when the thin film approaches the horizon,
the thickness
of the film vanishes and the entropy density expressed by Eq. (\ref{density}) doesn't change.
In this case, the thin film itself becomes horizon, the entropy of thin film 
is just the entroy of horizon, and the black hole entropy is just identified with the
entropy of the quantum field on the event horizon.

\section*{acknowledgments}

We would like to thank colleagues in our group for helpful discussion.
We are supported by the National Natural Science Foundation of China under
Grant No. 10073002.


\begin{references}

\bibitem{}J. D. Bekenstein, Phys. Rev. D {\bf 7}, 2333 (1973); {\bf 9}, 3292 (1974).
\bibitem{}S. W. Hawking, Nature (London) {\bf 248}, 30 (1974); Commun. Math. Phys.
          {\bf 43}, 199 (1975).
\bibitem{}G. W. Gibbons and S. W. Hawking, Phys. Rev. D {\bf 15}, 2752 (1977).
\bibitem{}G 't Hooft, Nucl. Phys. {\bf B256}, 727 (1985).
\bibitem{}Luo Zhi-jian and Zhu Jian-yang, Acta Phys. Sin. {\bf 48}, 395 (1999).
\bibitem{}Liu Wenbiao and Zhao Zheng, Chin. Phys. Lett. {\bf 18}, 310 (2001).
\bibitem{}Li Xiang and Zhao Zheng, Phys. Rev. D {\bf 62}, 104001 (2000).
\bibitem{}W. Kinnersley, Phys. Rev. {\bf 186}, 1335 (1969).
\bibitem{}Zhao Zheng and Dai Xian-Xin, Mod. Phys. Lett. A {\bf 7}, 1771 (1992).
\bibitem{}Zhao Zheng, Luo Zhi-qiang and Dai Xian-xin, Il Nuoro Cimento {\bf 109B}, 483 (1994).
\bibitem{}Zhao Zheng, Zhang Jian-hua and Jiang Ya-ling, Int. J. Theor. Phys. {\bf 36},
1359 (1997).
\bibitem{}T. Damour and R. Ruffini, Phys. Rev. D {\bf 14}, 332 (1976).
\bibitem{}Li Zhong-heng and Zhao Zheng, Acta Physica Sinica {\bf 46}, 1273 (1997).
\bibitem{}Min-Ho Lee and J. K. Kim, Phys. Rev. D {\bf 54}, 3904 (1996).
\bibitem{}Jeongwon Ho, Won T Kim, Young-Jai Park, and Hyeonjoon Shin, 
          Class. Quantum Grav. {\bf 14}, 2617 (1997).
\bibitem{}Li Xiang, Zhao Zheng, Mod. Phys. Lett. A {\bf 15}, 1739 (2000).
\bibitem{}R. M. Wald, {\it General Relativity}
          (The University of Chicago Press, Chicago, 1984).
\end{references}
\end{document}